\documentstyle[11pt,paspconf,epsf]{article}

\begin{document}

\title{Host galaxies and environment of BL Lac objects}
\author{J. Heidt}
\affil{Landessternwarte Heidelberg, K\"onigstuhl, 69117 Heidelberg, Germany}

\begin{abstract}
Since the last meeting on BL Lac objects 10 years ago, 
BL Lac host galaxies and their cluster environment have gained 
much attention. 
Hence, our current knowledge of the properties of BL Lac host galaxies
and their cluster environment has improved considerably, 
which will be reviewed. The importance of future observing programs 
using (very) large telescopes is briefly outlined.
\end{abstract}

\section{Introduction}

Since BL Lac objects are dominated by radiation from a (presumably)
relativistic jet pointing almost directly towards the observer, 
their host galaxies and  close environment are hard to study. 
This problem is accompanied by their cosmological distances. Naturally, its
a challenge for each observer in order to separate the contribution 
of the host galaxy or even the close environment from the BL Lac itself. 
For the cluster environment, the BL Lac itself is not a problem.
Here the observer has to deal with a wide range of uncertainties such
as the proper determination and separation of the background
from any putative cluster environment.
The study of both, the hosts and any cluster environments are 
hampered by the fact that large telescopes with sufficient resolution and 
large detectors (and good-will TACs) are needed in order to reach 
the limiting magnitudes required.

Still, such studies are important as they can provide answers for
at least some of the questions related to our understanding of BL Lac objects
and the AGN phenomenon in general. Most quoted questions are:
What are the hosts of BL Lac objects (disks/ellipticals), what is their
morphology and how do they compare to ``normal'' galaxies? Do they reside
in clusters? Do they show any cosmological evolution of their hosts or
cluster environment? Do their hosts or cluster environment show any 
correlation with intrinsic properties or differences between HBL and LBL?
How do the properties of the hosts and cluster environment compare to 
those of their ``parents'', the Fanaroff-Riley I (FR I) radio galaxies?
Are tidal forces important in triggering and maintaining the activity 
in these sources? And finally: Are (some) BL Lac objects actually lensed QSOs?

Although BL Lac objects are now known for about 30 years, much of our current 
understanding of the properties of their hosts
and environment has been emerged in recent years. This is mainly due to the
large telescopes offering excellent seeing conditions now available 
in combination with dedicated analysis packages. 

In this review I will first give a brief history about the first 20 years 
since the identification of BL Lac with a radio source. Than I will describe
step by step what we know about BL Lac host galaxies, their close environment
and their cluster environment and will finally present a few key issues,
which should be investigated in the future.

\section{History - the first 20 years}

Although the whole story begun already with the identification of 
the ``variable'' star BL Lac with a radio source by Schmitt (1968) $-$
where he noted ``..marginal nebulosity about the star.'' $-$
only few studies were carried out on this new class of objects.
Milestones in the first 
10 years were the (wrong) claim of a BL Lac in a cluster of galaxies 
(Disney, 1974), consistency of the surface brightness profile of BL Lac
with that of an elliptical galaxy at z = 0.07 (Kinman, 1975) and the first
spectroscopical confirmation of a close companion (30 kpc) to a BL Lac
by Baldwin et al. (1977). At the first BL Lac conference in Pittsburgh in 
about 8/40 BL Lac objects hosts were detected either
by aperture/annuli photometry and/or spectrophotometry (c.f. Miller, 1978).

Although the number of BL Lac objects grew by a factor of three during the
next decade, most observations still concentrated on the previously studied
objects making use of the new CCDs. Weistrop et al (1979) were the first, 
who carried out surface photometry using such a CCD
and decomposed the resulting 1-dim surface brightness profile into
core and host galaxy. Weistrop et al. (1981) noted that BL Lac objects
avoid clusters, and Halpern et al. (1986) were the first (wrongly) claiming
a BL Lac object to be hosted by a disk galaxy. At the second BL Lac conference
in Como the properties of the hosts of about 15/120 BL Lac objects were known
(Ulrich, 1989), but their close environment and cluster properties 
were still poorly understood.

The Como meeting essentially marked the ``kick-off''. With the upcoming
availability of large, well defined samples of BL Lac objects 
(e.g. the 1 Jy sample) systematic investigations were possible. 
Nowadays, $\sim$ 300 BL Lac objects are known, about one third have 
been imaged to study their hosts and about one quarter have been imaged for
their cluster environment. The main results of the last ten years will be
reviewed in the next sections.

\section{Host galaxies}

Systematical 1-dim/2-dim investigations of BL Lac host galaxies have 
been carried out 
e.g. by Abraham et al. (1991), Stickel et al. (1993), Wurtz et al. (1996), 
Falomo and references therein (1996), Heidt et al. (1998) in the optical,  
Kotilainen \& Falomo (1998) in the optical and NIR, and Heidt (1997) and
Wright et al. (1998) in the NIR. Basically the image of the object is 
decomposed into an AGN represented by either an observed or numerical PSF and 
a galaxy convolved with the PSF. The surface brightness of the galaxy 
can be modeled as:

\begin{displaymath}
{\rm I}({\rm r}) = {\rm I}_{\rm e} {\rm dex} \left\{ -{\rm b}_{\bf \beta} 
\left[ \left( \frac{{\rm r}}{{\rm r}_{\rm e}} \right) ^{\beta}
-1 \right]
\right\}
\end{displaymath}

where ${\rm r}_{\rm e}$ is the effective radius, ${\rm b}_{\beta}$ is a
$\beta$-dependent constant and $\beta$ the shape parameter (Caon et al. 1993).
A shape parameter $\beta$ = 0.25 represents a de Vaucouleurs profile 
(elliptical galaxy) and $\beta$ = 1 represents an exponential profile 
(disk galaxy). An example of a 2-dimensional decomposition is shown 
in Figure 1.
\begin{figure*}[t!]
\plottwo{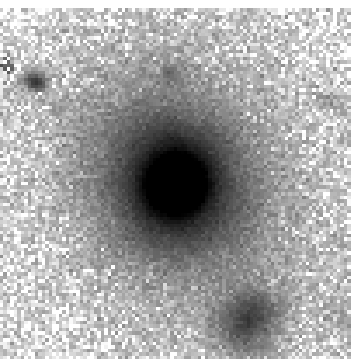}{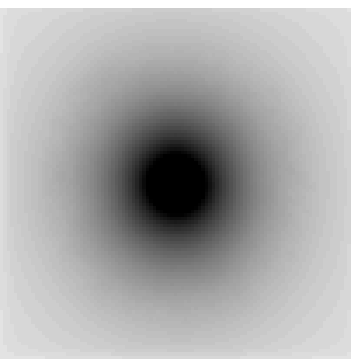}
\vspace*{.2cm}
\plottwo{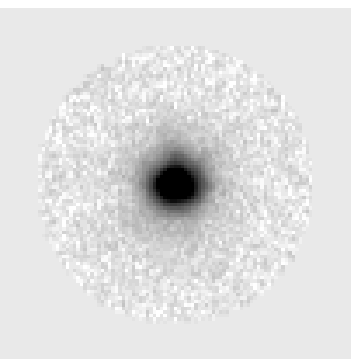}{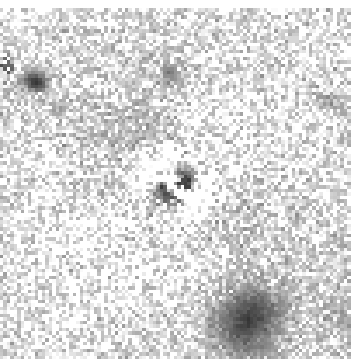}
\caption{2-dimensional decomposition of 1ES 1255+244 (z = 0.141). Top left)
Nordic Optical Telescope image of 1ES 1255+244 (18''$\times$18'', 
North is up, East to the left).
Top right) Convolved model galaxy. Bottom left) Scaled core (PSF).
Bottom right) Residuals after subtraction of convolved model galaxy and
scaled core. The residuals are less than 1\% of the total light. 
The best fit for the host galaxy is an elliptical galaxy with ${\rm M}_{\rm R}
= -23.2$ and ${\rm r}_{\rm e} =$ 2.1'' = 7.2 kpc.}
\end{figure*}
The host galaxies of BL Lac objects are luminous $({\rm M}_{\rm R} = -23.5\pm1$
and large ${\rm r}_{\rm e} = 10\pm$7 kpc 
elliptical galaxies\footnote{${\rm H}_0 = 50, {\rm q}_0 = 0$ assumed 
throughout}, i.e. they are $\sim$ 1 mag brighter than a 
${\rm L}^{\star}$ galaxy, but somewhat fainter than brightest cluster 
galaxies. On average, they are fainter than typical FR I hosts and
resemble rather typical FR II hosts. 

When leaving $\beta$ as free parameter, in most cases $\beta$ close 
to the canonical $\beta$ = 0.25 was found. Two of them may have inner disks
(PKS 0548-322 (Falomo et al. 1995), 1ES 1959+650 (Heidt et al. 1998)).
In all cases (except OJ 287 (Wurtz et al. 1996, Yanny et al. 1997)) the
host galaxy and the core are well centered on each other. They
show no correlation with intrinsic properties, no difference between the hosts
of HBL and LBL was found. The optical-NIR colours are typical for 
ellipticals with no evidence for recent star formation. Finally, they
show evidence for positive evolution with redshift. Unfortunately, only a 
few BL Lac host galaxies at redshifts above 0.7 have been studied. Hence,
the evidence for evolution is uncertain.

At the time of the conference, only few results on BL Lac hosts 
based on HST imaging have been published. The observations basically 
confirm ground-based results. The profiles follow even very close to the 
center (r$\leq$0.2'') the de Vaucouleurs law, the hosts and AGN are very 
well centered on each other (Falomo et al. 1997a, b,  Jannuzi et al. 1997).
For first results from the HST Snap survey on BL Lac objects providing 
information about $\sim$ 100 sources see Falomo et al. (these proceedings).

At least 4 BL Lac objects have been claimed/disproven to be hosted by
a disk-dominated system: MS 0205.7+3509, PKS 1413+135, 1E 1415.6+2557 
and OQ 530 (Halpern et al. 1986, Abraham et al. 1991, Stickel et al. 1993,
McHardy et al. 1991, 1994, Romanishin 1992, Stocke et al. 1992, 1995, 
Stickel et al. 1993, Wurtz et al. 1996, Falomo et al. 1997). In a few more 
according to the fits neither elliptical nor disk-type systems are preferred.
Most likely, the hosts of OQ 530 and PKS 1413+135 are disk-dominated systems.
However, PKS 1413+135 is a widely discussed object. It has a edge-on
early-type spiral, but may also be a lensed AGN. On the other hand, neither
off-centering between host and AGN nor multiple images have been detected in
this system.
 
There is only BL Lac object, which is clearly a lensed system (B2 0218+357,
Grundahl \& Hjorth, 1995).
There are  three more promising candidates being lensed systems,
investigated through both, imaging (claims of intervening galaxies) 
and spectroscopy (identification of intervening systems). 
These are AO 0235+164, PKS 0537-441
and B2 1308+326 (Falomo et al. 1992, 1996, 1997b, Stickel et al. 1993 and refs.
therein, Abraham et al. 1993, Nilsson et al. 1996, Burbidge et al. 1996). 
They all show similar properties, which makes them promising candidates:
I) they are at high redshift (z $\sim$ 1), II) they are among the most 
luminous BL Lac objects (${\rm M}_{\rm R} < -$27.5) and III) they have 
historically shown the largest variability amplitudes 
($\Delta$m $\sim$ 5 mag) of all BL Lac objects. 
However, even during the course of a long-term project aiming in 
detecting microlensing events in a large sample of QSOs and BL Lac objects,
no clear microlensing event was detected (Borgeest \& Schramm 1993). 
Signatures for lensing (e.g. multiple images) has only
been observed in B2 0218+357. 

Further candidates are MS 0205.7+3559 and PKS 1413+135 and virtually
all 7 BL Lac objects, where signatures of absorbing systems in their
spectra have been observed. In fact, in two of those (PKS 0139-097 and 
PKS 0118-272) absorbing systems have been detected through imaging 
(Heidt et al. 1996, Vladilo et al. 1997).

Still, based on the small number of potential candidates and given the
amount of data available lensing effects are unlikely to be 
important to the BL Lac phenomenon.

\section{Local environment}

Contrary to the host galaxies, the local environment (projected distance 
$<$ 100 kpc from the BL Lac) is as yet poor studied due to the faintness of 
most neighboring galaxies (3-5 mag fainter as the BL Lac). Therefore,
most information have been derived from deep high-resolution imaging.

Evidence for interaction (projected distance 25-100 kpc) with 
bright companions have
been observed through imaging and partly been confirmed via spectroscopy
e.g. in  PKS 0548$-$322, Ap Lib and 3C 371 (Stickel et al. 1993, Pesce et al. 
1994, Falomo et al. 1995, Nilsson et al. 1997). Recently, three more very good
candidates with bright companions as close as 10 kpc (among them one merger 
remnant?) have been found (1ES 1440+122, 1ES 1741+196 and 1ES 1853+671
(Heidt et al. 1998a, b)). The case of 1ES 1741+196 is shown in Figure 2.

\begin{figure}[t!]
\plottwo{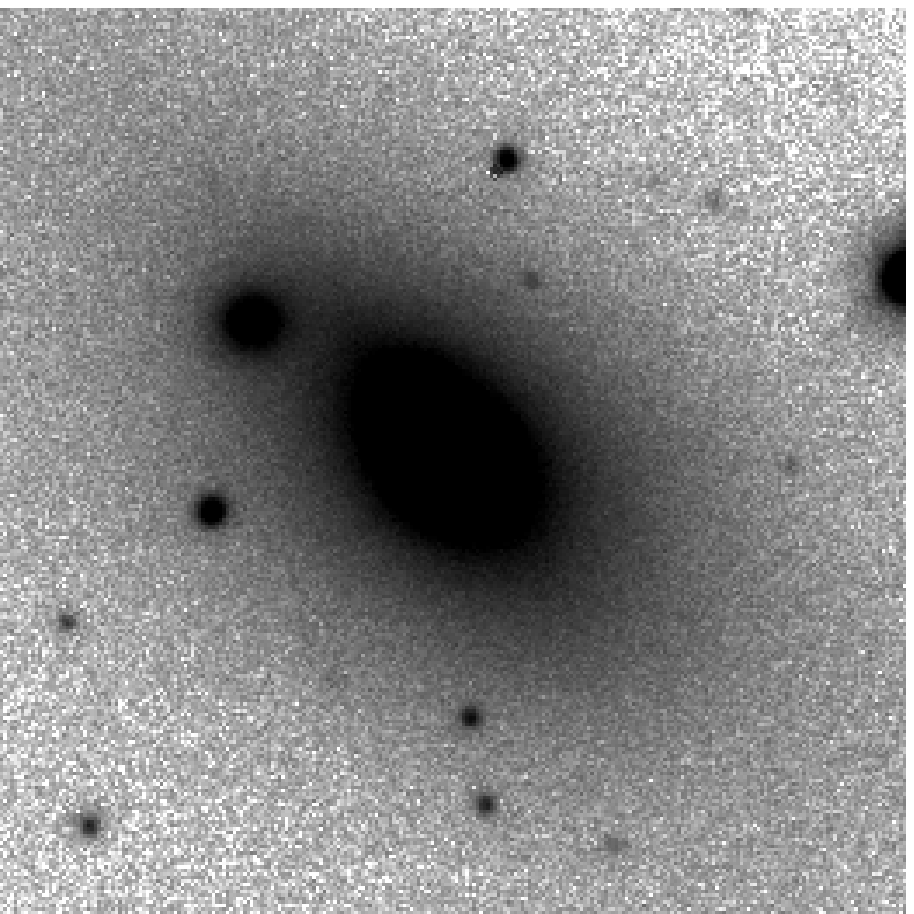}{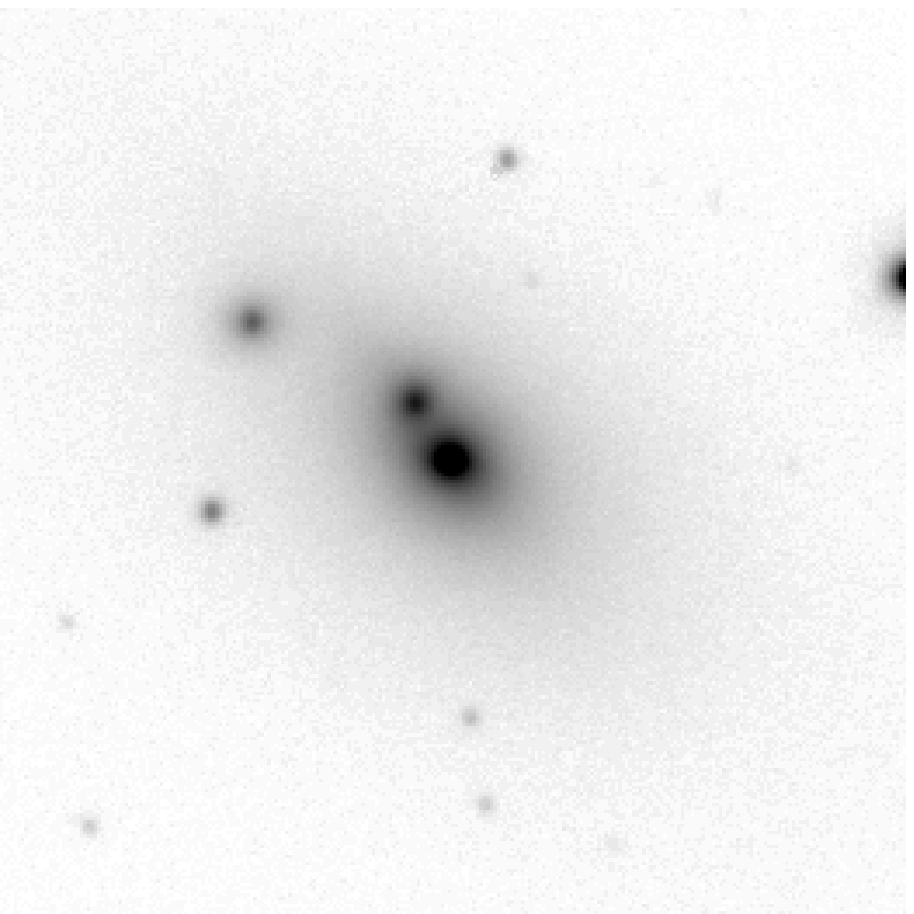}
\vspace*{.2cm}
\plottwo{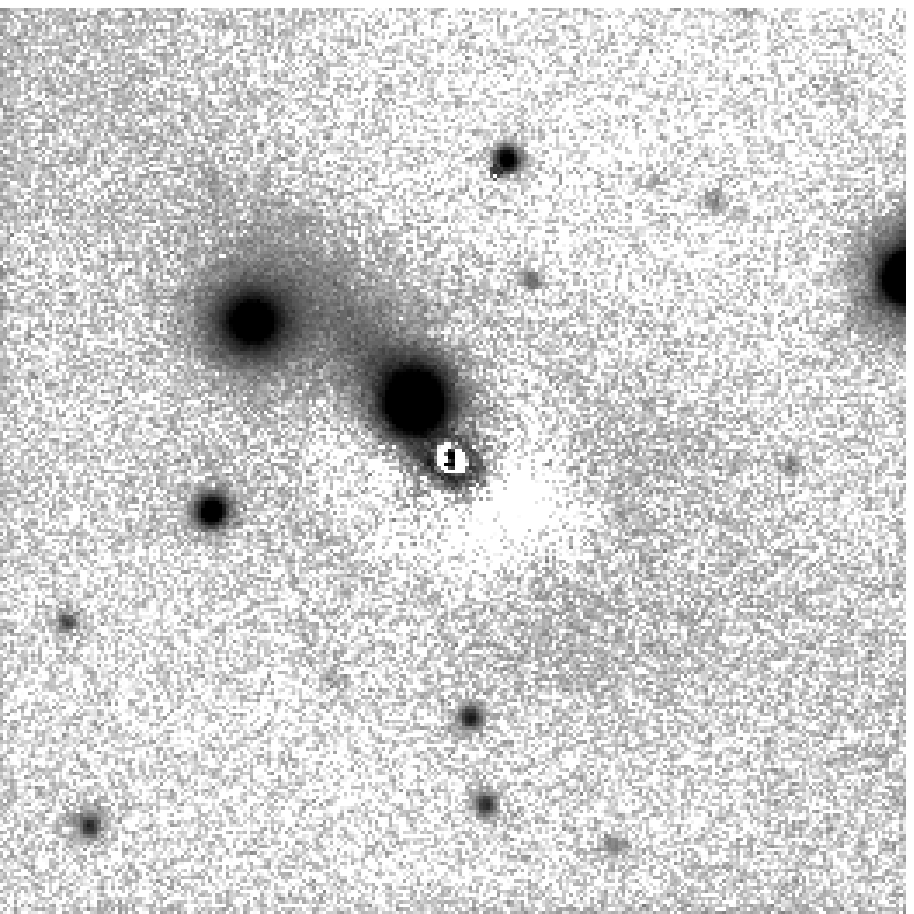}{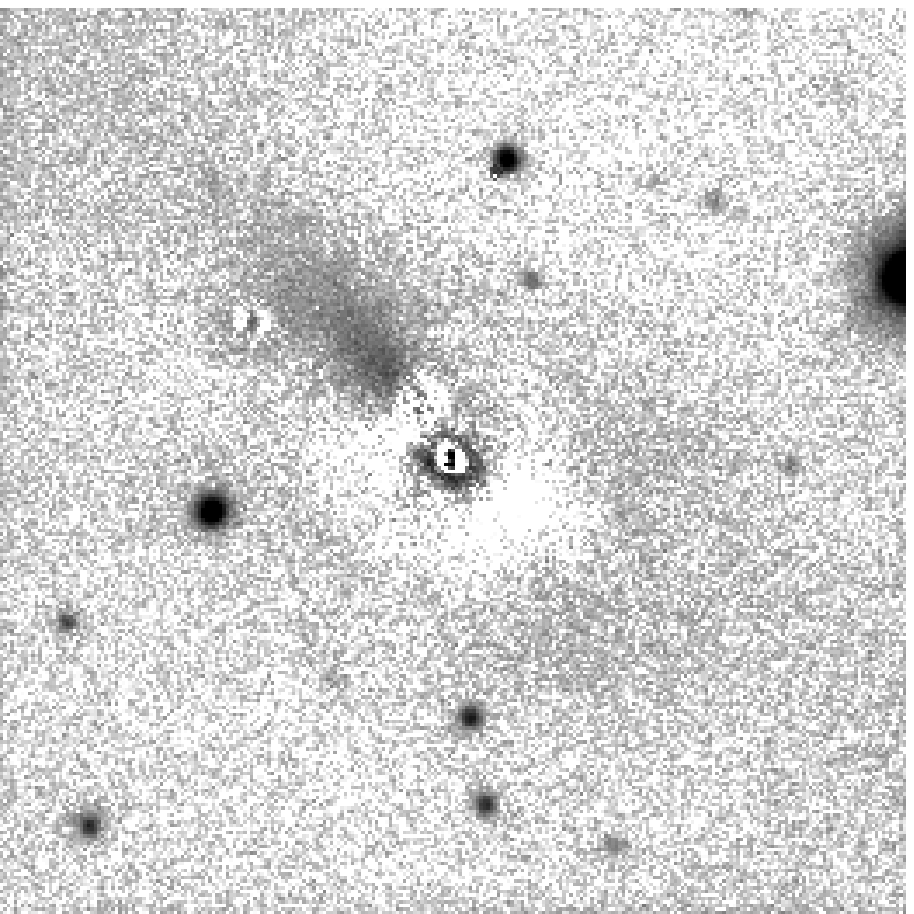}
\caption{Top) Nordic Optical Telescope image of 1ES 1741+196 (z = 0.083)
shown with two different dynamic ranges in order to show the extent of the
host galaxy of 1ES 1741+196 (left) and two closeby companions (right).
Field is 44'' $\times$ 44'', North is up, East to the left.
Bottom left) Same image after subtraction of the model for the host galaxy
of 1ES 1741+196 and the core. The host galaxy is a very luminous, large 
elliptical galaxy (${\rm M}_{\rm R} = -24.4$, ${\rm r}_{\rm e} = 25$ kpc).
The two closeby companions are clearly visible. They have the
same redshift as the BL Lac and are at projected distances of
7.3 and 27 kpc, respectively. Thus it  is most likely a triplet of currently 
interacting galaxies. Note the tidal bridge connecting the
companion galaxies to 1ES 1741+196. Bottom right) Same image after 
subtraction of models for the two closeby companion galaxies to show the
tidal bridge more clearly. The closest companion  
(7.3 kpc) is a disk-type galaxy with ${\rm M}_{\rm R} = -21.3$, the more 
distant companion an elliptical galaxy with ${\rm M}_{\rm R} = -20.4$.
It is not clear, if the tidal bridge continues towards the BL Lac.}
\end{figure}

\begin{figure}[t!]
\caption{Nordic Optical Telescope image of the large scale environment 
of 1ES 1255+244 (z = 0.141). The BL Lac (in the center)  is 
located in a loose group of galaxies.
Field is 3'$\times$3' (610 kpc $\times$ 610 kpc at z = 0.141). North is up,
East to the left.}
\end{figure}

In most BL Lac objects, companion galaxies (without obvious signs of 
interaction) have been detected within 100 kpc radius, but their physical 
association has been confirmed only for some of the brighter companions 
galaxies (e.g. Falomo et al. 1991, 1993, 1995, Romanishin, 1992, 
Stickel et al. 1993, Pesce et al. 1994, 1995). Even more close 
(projected distance $<$50 kpc) , (mostly) faint companions have also been 
detected for a substantial number of BL Lac objects from ground (e.g. Falomo et al. 
1990, 1996, Benitez et al. 1996, Heidt et al. 1998a) and with HST
(Falomo et al. 1997a, b, Yanny et al, 1997, Jannuzi et al. 1997).

The high frequency of close companion galaxies and the evidence for
gravitational interaction in some sources  indicates that tidal forces
might be important for triggering and maintaining the activity in at least
some BL Lac objects. Based on the small amount of data available it is not 
clear, if this is a general requirement. The main problem is the lack of
redshifts for most of the companion galaxies. A systematic study
of this phenomenon and the comparison with ``normal'' galaxies,
might be an important work in the future.

\section{Cluster environment}

Characterizing and quantifying the cluster environment of BL Lac objects
is a rather tricky subject. Since spectroscopical studies of a large number
of galaxies nearby BL Lac objects are entirely lacking (and unlikely to be
conducted in the very near future), the analysis is based on the examination
of an excess of galaxies in the fields around BL Lac objects as 
compared to the background. This is hampered by several problems,
e.g. the proper determination of the background (and foreground for
high-redshift sources) due to the limited field of view, limiting
magnitudes of the observations, seeing etc. 

There are two principles to estimate an excess of galaxies within a given
radius around the BL Lac (typically 500 kpc) as compared to the
background. The first is the calculation of the spatial covariance amplitude 
as introduced by Longair \& Seldner (1979), the second, more robust estimate,
is to count the number of galaxies above a certain magnitude limit
determined by the third-ranked cluster galaxy 
(m $< {\rm m}_{\rm 3} + 2$, where ${\rm m}_{\rm 3}$ is the third-ranked
cluster galaxy (Bahcall 1981)). For both methods, the resulting numbers 
can than be converted to Abell richness classes.

Systematic studies of the cluster environment have been carried out
e.g. by Fried et al. (1993), Pesce et al. (1993, 1994), Smith et al. (1995)
and Wurtz et al. (1993, 1997). Although the results obtained by the authors
differ for some objects, the main conclusions are that most BL Lac objects are
located in poor clusters (Abell $\leq$ 0). There is an indication for
evolution with redshift (richer clusters) and there is no good match with
the cluster environment of FR I. However, the properties of the cluster 
environment at redshifts above 0.7 are completely unknown. An example of a
typical cluster environment is shown in Figure 3.

\section{Summary}

Our current knowledge of the host galaxies and cluster environment
of BL Lac objects can be summarized as follows:

1) BL Lac objects are hosted by luminous and large elliptical galaxies
(${\rm M}_{\rm R} = -$23.5, ${\rm r}_{\rm e}$ = 10 kpc). 
2) Many of them have closeby
($<$ 50 kpc projected distance) companion galaxies, some of them show
evidence for (recent) interaction. 3) They are located in poor galaxy
clusters (Abell $\leq$ 0). 4) There is no correlation of the
hosts or cluster environment with intrinsic properties and no difference
of the hosts or cluster environment between HBL and LBL.
5) Both, the host galaxies and cluster environment show indication
for evolution with redshift (brighter hosts and denser environments).
6) The properties of the hosts and cluster environment of BL Lac objects 
do not match those of the FR I very well. Therefore the Unified Scheme
has to be modified. Either, FR II should at least in part be included,
or the brightest FR I and/or the FR I in rich clusters have to be excluded.

\section{Future work}

Since we are now know the general properties of the hosts and cluster 
environment of (at least nearby) BL Lac objects quite well,
we can now start to tackle a few issues in more detail.
Currently, there are several projects underway (c.f. Laurent-Muehleisen this 
proceedings), which will strongly increase 
the number of BL Lac objects thus allowing to construct well defined samples
of BL Lac objects e.g. at low redshifts. From the observational side,
the data gathered during the HST Snap Survey offer a unique opportunity to
study the host galaxies of BL Lac objects with exceptional resolution and
to identify close and faint companion galaxies. Very large telescopes
(VLT, HET, LBT) will allow to carry out spectroscopy of BL Lac host galaxies
and even very faint companion galaxies. Future observing programmes 
should include:

\begin{itemize}
\item Imaging of high-redshift BL Lac objects (perhaps also those without known 
redshift). Can the trend of evolution for both, the host galaxies and 
cluster environment be confirmed? This would than also allow a detailed
comparison with other types of radio-loud AGN.

\item Multicolour high-resolution imaging of the hosts of low-redshift 
BL Lac objects. Detailed analysis of the host properties (colors, faint disks,
ellipticities, disky/boxy-type ellipticals via Fourier analysis etc.)
to determine the impact of the core on the host and vice versa, the influence
of the environment and comparison with ``normal'' galaxies.

\item High-resolution imaging and spectroscopy of representative 
low-redshift samples of BL Lac objects, FR I, FR II and ``normal'' galaxies
to determine, if tidal forces (either through close companion galaxies or
gravitational interaction) are a requirement for triggering and
maintaining the activity in BL Lac objects. This is a tricky work,
in order to avoid possible biases the samples must be chosen very carefully
(e.g. similar distribution in luminosity, redshift and 
large-scale environment).
\end{itemize}

\acknowledgments I thank the organizers of the meeting for the kind
invitation for this review and for finanzial support.
This work was supported by the Deutsche Forschungsgemeinschaft (SFB 328).


\begin{references}
\reference Abraham, R.G., M$^{\rm c}$Hardy, \& I.M., Crawford, C.S. 1991, 
\mnras, 252, 482
\reference Abraham, R.G., Crawford, C.S., Merrifield, M.R., Hutchings, J.B., 
\& M$^{\rm c}$Hardy, I.M. 1993, \apj, 415, 101
\reference Bahcall, N.A. 1981, \apj, 247, 787
\reference Baldwin, J.A., Wampler, E.J., Burbidge, E.M., O'Dell, S.L.,
Smith, H.E., Hazard, C., Nordsieck, K.H., Pooley, G., \& Stein, W.A. 1977, 
\apj, 215, 408
\reference Ben\'{i}tex, E., Dultzin-Hacyan, E., Heidt, J., Sillanp\"a\"a, A.,
Nilsson, K., Pursimo, T., Teerikorpi, P., \& Takalo, L.O. 1996, \apjl, 464, L47 
\reference Borgeest, U., \& Schramm, K.-J., 1994, \aap, 284, 764 
\reference Burbidge, E.M., Beaver, E.A., Cohen, R.D., Junkkarinen, V.T.,
\& Lyons, R.W. 1996, \aj, 112, 2533  
\reference Caon N., Capaccioli M., \& D'Onofrio M. 1993, \mnras, 265, 1013 
\reference Disney, M.J. 1974, \apjl, 193, L103
\reference Falomo, R., Melnick, J., \& Tanzi, E.G. 1990, Nature, 345, 692
\reference Falomo, R., Giraud, E., Melnick, J., Maraschi, L., Tanzi, E.G.,
\& Treves, A. 1991, \apjl, 380, L67
\reference Falomo, R., Melnick, J., \& Tanzi, E.G. 1992, \aap, 255, L17
\reference Falomo, R., Pesce, J.E., \& Treves, A. 1993, \apjl, 411, L63
\reference Falomo, R., Pesce, J.P., \& Treves, A. 1995, \apjl, 438, L9 
\reference Falomo, R. 1996, \mnras, 283, 241 
\reference Falomo, R., Urry, C.M., Pesce, J.P., Scarpa, R., Giavalisco, M.,
\& Treves, A. 1997a, \apj, 476, 113
\reference Falomo, R., Urry, C.M., Pesce, J.P., Scarpa, R., Treves, A., 
\& Giavalisco, M. 1997b, Proc. ESO Conf. on Quasar Hosts, ed. D.L. Clements, 
I. P\'{e}rez-Fournon (Berlin:Spinger) 
\reference Falomo, R., Kotilainen, J., Pursimo, T., Sillanp\"a\"a, A.,
Takalo, L.O., \& Heidt, J. 1997c, \aap, 321, 374 
\reference Fried, J.W., Stickel, M., \& K\"uhr, H. 1993, \aap, 268, 53
\reference Grundahl, F., \& Hjorth, J. 1995, \mnras, 275, L67
\reference Halpern, J.P., Impey, C.D., Bothun, G.D., Tapia, S., Skillman, E.D.,
Wilson, A.S., \& Meurs, E.J.A. 1986, \apj, 302, 711 
\reference Heidt, J., Nilsson, K., Pursimo, T., Takalo, L.O., \& 
Sillanp\"a\"a, A.1996, \aap, 312, L13
\reference Heidt, J. 1997 Proc. ESO Conf. on Quasar Hosts, ed. D.L. Clements, 
I. P\'{e}rez-Fournon (Berlin:Spinger) 
\reference Heidt, J., Nilsson, K., Sillanp\"a\"a, A., Takalo, L.O., 
\& Pursimo, T. 1998a, \aap, in press
\reference Heidt, J., Nilsson, K., Fried, J.W., Sillanp\"a\"a, A., 
Takalo, L.O., \& Pursimo, T. 1998b, \aap, submitted
\reference Jannuzi, B.T., Yanny, B., \& Impey, C. 1997, \apj, 491, 146
\reference Kinman, T.D., 1975, \apjl, 197, L49
\reference Kotilainen, J., Falomo, R., \& Scarpa, R. 1998, \aap, 336, 479
\reference Longair, M.S., \& Seldner, M. 1979, \mnras, 189, 433 
\reference M$^{\rm c}$Hardy, I.M., Abraham, R.G., Crawford, C.S., 
Ulrich, M.-H., Mock, P.C., \& Vanderspeck 1991, \mnras, 249, 742
\reference M$^{\rm c}$Hardy, I.M., Merrifield, M.R., Abraham, R.G., \& 
Crawford, C.S. 1994, \mnras, 268, 681
\reference Miller, H.R. 1978, Proc. Pittsburgh Conference on BL Lac Objects,
ed. C.M. Wolfe (University of Pittsburgh)
\reference Nilsson, K., Charles, P.A., Pursimo, T., Takalo, L.O., 
Sillanp\"a\"a, A., \& Teerikorpi, P. 1996, \aap, 314, 754 
\reference Nilsson, K., Heidt, J., Pursimo, T., Sillanp\"a\"a, A., 
Takalo, L.O., \& J\"ager, K. 1997, \apjl, 484, L107
\reference Pesce, J.P., Falomo, R., \& Treves, A. 1994, \aj, 107, 494
\reference Pesce, J.P., Falomo, R., \& Treves, A. 1994, \aj, 110, 1554
\reference Romanishin, W. 1992, \apjl, 401, L65
\reference Schmitt, J.L. 1968, Nature, 218, 663
\reference Smith, E.P., O'Dea, C.P., \& Baum, S.A. 1995, \apj, 441, 113
\reference Stickel, M., Fried, J.W., \& K\"uhr, H. 1993, \aaps, 98, 393  
\reference Stocke, J.T., Wurtz, R.E., Wang, Q., Elston, R., \& Jannuzi, B.T.
1992, \apjl, 400, L17
\reference Stocke, J.T., Wurtz, R.E., \& Perlman, E.S. 1995, \apj, 454, 55 
\reference Ulrich, M-H. 1989, Proc. BL Lac Objects, ed. L. Maraschi, 
T. Maccacaro, M.-H. Ulrich (Berlin:Springer)
\reference Vladilo, G., Centurion, M., Falomo, R., \& Molaro, P. 1997,
\aap, 321, 45
\reference Weistrop, D., Smith, B.A., \& Reitsema, H.J. 1979, \apj, 233, 504
\reference Weistrop, D., Shaffer, D.B., Mushotzky, R.F., Reitsema, H.J., \& 
Smith, B.A. 1981, \apj, 249, 3
\reference Wright, S.C., M$^{\rm c}$Hardy, I.M., Abraham, R.G., 
\& Crawford, C.S. 1998, \mnras, 296, 961 
\reference Wurtz, R.E., Stocke, J.T., \& Yee, H.K.C. 1993, \aj, 106, 869
\reference Wurtz, R.E., Stocke, J.T., \& Yee, H.K.C. 1996, \apjs, 103, 109
\reference Wurtz, R.E., Stocke J.T., Ellingson, E., \& Yee, H.K.C. 1997,
\apj, 480, 547
\reference Yanny, B., Jannuzi, B.T., \& Impey, C. 1997, \apjl, 484, L113


\end{references}
\end{document}